\documentclass[amsmath,amssymb,groupedaddress,superscriptaddress,lengthcheck,aps,prb] {revtex4-1}
\usepackage{graphicx}
\begin{document}
\title{Neutral Collective Modes in Spin-Polarized Fractional Quantum Hall States at Filling Factors 1/3, 2/5, 3/7, and 4/9 }
\author{Dwipesh Majumder}
    \affiliation{Department of Physics, Santipur College, Nadia 741 404, India}
    \affiliation{Department of Theoretical Physics, Indian Association for the Cultivation of Science, Jadavpur, Kolkata 700 032, India}

\author{Sudhansu S. Mandal}
    \affiliation{Department of Theoretical Physics, Indian Association for the Cultivation of Science, Jadavpur, Kolkata 700 032, India}

\date{\today}

\begin{abstract}
 We determine the lowest and higher order collective modes in both spin-conserving and spin-reversed sectors by calculating energy differences of
the appropriate linear combinations of different levels of composite-fermion-excitons and the fully spin-polarized ground states at filling factors 
$\nu=1/3$, 2/5, 3/7, and 4/9. Apart from providing the detailed study of previously reported modes that have also been
observed in the experiments, we predict additional higher energy modes at different filling factors. The lowest and the next higher
spin-conserving modes have equal number of ``magneto-rotons'' and the number is the same as the number of filled effective Landau-like levels of
composite fermions. The higher
energy modes at $\nu =1/3$ merge with the lowest mode at long-wavelength. The spin-conserving modes do not merge at other filling factors.
Apart from showing zero-energy spin-wave mode at zero momentum, thanks to Larmor's theorem, the lowest spin-reversed modes at the 
ferromagnetic ground states of $\nu=2/5$, 3/7, and 4/9 display one or more ``spin-rotons'' at negative energies signalling the unstable
fully polarized ground states at sufficiently small Zeeman energies. The high energy spin-reversed modes also have spin-rotons but at 
positive energies.
The energies of these excitations depend on the finite width of the quantum well as the Coulomb interaction gets screened. We determine
finite thickness correction to the Coulomb interaction by the standard method of local density approximation and use them to calculate the 
critical energies such as rotons, long-wavelength, and short-wavelength 
modes which are detectable in inelastic light scattering experiments. 
\end{abstract}


\maketitle


\section{Introduction}

Electrons restricted into the Hilbert space of the partially filled lowest Landau level (LLL) by the application of strong magnetic field 
 perpendicular to the plane of the
quasi-two dimensional systems become strongly correlated that leads to a topological quantum state of matter giving rise to
fractional quantum Hall effect  \cite{Tsui82,Laughlin83} (FQHE). One of the topological characters appears through the emergence of composite fermions \cite{Jain89}
(CFs) from the cooperative phenomenon between electrons confined in the LLL. A composite fermion is a bound state of an electron and an even number $2p$ of quantized
vortices that produce a Berry phase of $2\pi$ for a closed loop around it. Because of this, CFs experience a much reduced magnetic field
$B^* = B_\perp - 2pn_e\phi_0$, where $B_\perp$ is the perpendicular component of the external magnetic field, $n_e$ is the two-dimensional
electron density, and the unit flux quantum  $\phi_0 = hc/e$. As a consequence, CFs form their own Landau-like kinetic energy levels, called $\Lambda$
levels ($\Lambda$Ls), in this reduced effective magnetic field. This energy splitting  in the LLL is thus a direct manifestation
of the topological order in FQHE. Integer quantum Hall effect (IQHE) of noninteracting CFs at filling factor $\nu^*=n$ (the integer $n$ refers to the
$n$ completely filled $\Lambda$Ls) describes FQHE of electrons at the filling factor $\nu = n/(2pn\pm 1)$, where $+(-)$ 
denotes $B^* >0$ $(<0)$. Jain's composite fermion wave functions \cite{Jain89} for the 
ground state of FQHE at $\nu = n/(2np+1)$ is
$\Psi_{\frac{n}{2np+1}} (z_1,\cdots,z_N) = P_{{\rm  LLL}} \prod_{i<j}^N (z_i - z_j)^{2p} \Phi_n (z_1,\cdots,z_N)$, where $z_j = x_j - iy_j$ is the complex coordinate of the $j^{{\rm th}}$ particle in $N$ particle
system, the Jastrow factor    $\prod_{i<j}^N (z_i - z_j)^{2p}$ represents the attachment of $2p$ quantized vortices to every electrons transforming
them into CFs, $\Phi_n$ is the wave function for completely filled $n$ lowest $\Lambda$Ls by noninteracting CFs, and $P_{{\rm  LLL}}$ represents
projection into the LLL. These wave functions describe the wave functions for electrons being spinless which is justified 
because the electrons become polarized for large Zeeman energy $E_z = \vert g \vert
\mu_B B$, where $g$ is the effective g-factor for the electrons in the system and $B$ is the total external magnetic field. Nevertheless, the polarization
of the FQHE states can become less than $100\%$ for low Zeeman energy, 
to be precise, polarization \cite{Park98} will be $(n_{\uparrow}-n_{\downarrow})/(n_\uparrow + n_\downarrow)$ where $n=n_\uparrow + n_\downarrow$ and
 $n_\uparrow (n_\downarrow)$ represents 
the number of $\Lambda$Ls filled by up (down)-spin CFs, and consequently spin-transitions \cite{Du95,Eisenstein90,Kukushkin99,Yacoby07}take place by tuning $B$ and keeping $B_\perp$ fixed.
In this paper we consider fully spin-polarized FQHE states only, but the role \cite{Mandal01} of empty spin-reversed $\Lambda$Ls on neutral collective excitations will
also be considered.

Neutral collective modes in quantum Hall states have been studied using Hartree-Fock approximations \cite{Kallin84,Longo93} for the electronic excitons, 
density modulation \cite{Feynman} over the ground state in the single-mode approximation \cite{Girvin85,Park00} (SMA), exact diagonalizations 
\cite{Platzman94}for small systems, Hamiltonian description of composite fermions \cite{Murthy99}, and excitons of composite fermions \cite{Dev92,Scarola00,Mandal01}.
All these studies qualitatively or semiquantatively describe the presence of ``magneto-roton'' and ``spin-roton'' in the collective modes that have been
identified in several inelastic light scattering (ILS) experiments.\cite{Pinczuk93,Kang01,Dujovne03,Dujovne05,Majumder11_2,Majumder11_1}
In this paper, we employ the method of using excitons of CFs for determining neutral collective modes.
In the composite-fermion theory, the neutral collective excitations emerge due to mixing \cite{Majumder09,Majumder11_2,Majumder11_1}
of several levels of CF-excitons, wherein a single CF from any of the filled
$\Lambda$Ls is excited into any of the empty $\Lambda$Ls of either spin. If the change in the $\Lambda$L index of the excited CF becomes $\lambda$,
then the corresponding CF-exciton is called ``level-$\lambda$'' CF-exciton. Apart from $\nu=1/3$, there can be more than one CF-excitons at
level-$\lambda$, depending on the $\Lambda$L where the said CF is situated in the ground state. The excitation of a CF into a $\Lambda$L without (with)
changing its spin, {\it i.e}, the change in spin projection $\vert \Delta s_z \vert = 0\, (1)$ corresponds to spin-0 (spin-1) CF-exciton. Therefore,
the excited state wave function \cite{Dev92,Wu96} that corresponds to level-$\lambda$ and spin-$s$ ($s$=0 or 1) CF-exciton is $\Psi^{{\rm ex}} (z_1,\cdots,z_N)
= P_{{\rm  LLL}} \prod_{i<j}^N(z_i-z_j)^{2p} \chi_{n,\lambda}^s (z_1,\cdots,z_N)$ where $\chi_{n,\lambda}^s$ represents wave function
of a CF-exciton where one of the CF from $n$-filled $\Lambda$Ls is excited across $\lambda$ $\Lambda$Ls with  $\vert \Delta s_z \vert =s$. 
This theory naturally describes more than one neutral collective modes \cite{Majumder09,Majumder11_2,Majumder11_1} and thus it is clearly distinct from the 
SMA which by definition describes only one neutral mode. 
In past, both spin-conserving (spin-zero) and spin-reversed (spin-one) neutral collective modes had been studied \cite{Girvin85,Park00,Longo93} for several filling 
factors using SMA. While some of the qualitative features such as number of ``magneto-roton'' minima and finite gap at the long wavelength limit in
spin-zero modes and spin-wave mode in the filling factor 1/3 agree with the lowest collective modes in the CF theory, the latter describes following
additional characteristics which do not have any SMA analogue. First, the higher energy spin-zero modes at $\nu=1/3$ merge \cite{Majumder09} with the lowest
mode at the long wavelength limit.
Second, the higher energy spin-zero modes at $\nu =1/3$ and 2/5, and spin-one modes at $\nu=1/3$ also have rotons in their dispersions \cite{Majumder09,Majumder11_1}.
The lowest spin-one modes at $\nu=2/5$, 3/7, and 4/9 have ``spin-roton'' minima \cite{Majumder11_2} showing an unusual characteristic of a 
composite fermion ferromagnet as their energies are lower than $E_z$. 
In this paper, we provide a detailed account of the theoretical aspects of our previously published results \cite{Majumder09,Majumder11_1,Majumder11_2},
and a handful of new results: We determine the lowest two spin-zero collective modes at filling factors 3/7 and 4/9, and
spin-one collective modes at higher energies for filling factors 2/5, 3/7, and 4/9. By incorporating local density approximation \cite{Ortalano97,Meshkini} (LDA), 
we calculate quantum well thickness and carrier density dependent energies of the critical modes such as rotons, spin-rotons, 
long-wavelength, and large-momentum for all of the neutral collective modes.

   The rest of this paper is organized as follows. In the next section, we provide explicit form of the composite fermion wave functions for the ground
states as well as excited states due to the formation of both spin-0 and spin-1 CF-excitons at 
filling factor $\nu = n/(2n+1)$ in a spherical geometry\cite{Haldane}. We have derived a recursion relation for the Jain-Kamilla \cite{JK} LLL-projected spherical
harmonics as the single particle basis states. In section III, we present both spin-zero and spin-one collective modes obtained by
the evaluation of Coulomb energies for the composite fermion wave functions of the ground and excited states by the Monte Carlo method in the fully
polarized phase of the filling factors 1/3, 2/5, 3/7, and 4/9. First, we obtain wave functions for different levels of CF-excitons when a CF particle is 
excited across some of the lowest available empty $\Lambda$Ls with or without same spin, depending on the type of excitations. Secondly, we calculate
all the elements of the Coulomb matrix in the restricted low-energy Hilbert space of the chosen levels of CF-excitons. We then 
perform Graham-Schmidt orthogonalization procedure to obtain an orthogonal basis in this restricted Hilbert space and calculate effective Coulomb
matrix in the new orthogonal basis. Finally, we diagonalize the effective Coulomb matrix \cite{MJ02} to obtain the energies of the excited states and hence 
the determination of the energies of the lowest and higher modes upon subtracting the energy of the fully polarized ground state. The characteristics
of the modes at different filling factors are as follows. Filling factor 1/3:
All the lowest three spin-zero modes that we determine have one roton and they merge \cite{Majumder09}at the long wavelength limit, although
they are well separated at the high momentum region. The lowest spin-one mode is a spin-wave with zero interaction energy at the momentum $q=0$, in
consistence with Larmor's theorem, and finite energy at large momentum. The next two higher spin-one modes are well separated with the formation of one
spin-roton \cite{Majumder11_1}in each of those. Filling factor 2/5: The lowest two spin-zero modes have two rotons each and the modes do not merge 
\cite{Majumder09}at long wavelength. The
lowest spin-one mode behaves as spin-wave at long-wavelength but its negative curvature leads to lowering its energy until it forms a spin-roton \cite{Majumder11_2} 
minimum; in a window of momentum, the energy of this mode is lower than the Zeeman energy. The next higher spin-one mode also possesses a spin-roton
but it has finite energy at $q=0$. Filling factor 3/7: The lowest two spin-zero modes have three rotons each and finite separation at $q=0$.
As in filling factor 2/5, the lowest spin-one mode behaves as spin-wave at long wavelength, forms a spin-roton, and shows negative energy \cite{Majumder11_2} in a window
of momentum. The next higher spin-one mode has one spin-roton and has finite energy at $q=0$. Filling factor 4/9: The lowest two spin-zero modes
have four rotons each. The lowest spin-one mode has two spin-rotons and a spin-maxon (a maximun in the dispersion of spin-one mode),
apart from its spin-wave nature at $q=0$. The net excitation energy of
this mode is always less than $E_z$. The next higher mode has a spin-roton.
In section IV, we review the procedure \cite{Ortalano97,Meshkini} of determining effective two dimensional Coulomb potential between electrons in a square well of finite transverse width
in LDA. We use this effective potential to determine all the modes discussed above. The dependence of the 
critical energies of the modes, {\it viz.}, the rotons, spin-rotons, long wavelength, and short wavelength on the width of the quantum wells and electron
densities are obtained for the filling factors 1/3, 2/5, 3/7, and 4/9. Section V is devoted for conclusion where we discuss about the experimental realization
of our findings.

\section{Composite Fermion Wave Function in Spherical Geometry}

We employ standard spherical surface \cite{Haldane} where the electrons are influenced by the radial magnetic flux $2Q \Phi_0$ ($2Q$ is an integer)
due to the Dirac magnetic monopole of charge $Q$ placed at the center of the sphere with radius $R= \sqrt{Q}\ell$, where $\ell = (\hbar c /eB_\perp)^{1/2}$ is
the magnetic length. The single particle wavefunctions in such a geometry 
are the spherical harmonics \cite{Yang}: 
\begin{eqnarray}
&&Y_{Q,l,m}(\Omega)= N_{Qlm} e^{iQ\phi} u^{Q+m} v^{Q-m} \\ \nonumber
&\times&\sum_{s=0}^{l}(-1)^s {{l \choose s}} {{ 2Q+l \choose Q+l-m-s}}
 (v^*v)^{l-s}(u^*u)^s\;\;
\label{single_part}
\end{eqnarray}
where $\Omega$ represents the collective spherical spinor variables
$u=\cos(\theta/2)\exp (-i\phi/2)$ and $v=\sin(\theta/2)\exp (i\phi/2)$  with $0 \leq \theta \leq \pi$ and $0 \leq \phi < 2\pi$,
$l=0,1,2,\cdots$ denote the energy levels known as Landau levels, the degenerate states labeled by
$m=-(Q+l), -(Q+l)+1, \dots , (Q+l)-1, Q+l$ for 
 $l$-th LL, and $ N_{Qlm}$ is the normalization constant. 
The IQHE wave function for filling factor $\nu = n$ is the Slater determinant corresponding to the lowest $n$-filled Landau levels (shells), and can
be represented by $\Phi_n(\Omega_1,\cdots,\Omega_N)$ for $N$ noninteracting electrons.

\subsection{Ground state wave function}

In the presence of repulsive interaction between electrons, CFs are formed and the reduced magnetic flux experienced by them in the
spherical geometry is $2q = 2Q -2p(N-1)$ for a system of $N$ electrons. A set of effective Landau-like levels, known as $\Lambda$Ls, are formed for noninteracting
CFs with their reduced flux. IQHE of such CFs at effective filling factor $\nu^* = n$ corresponds to FQHE of electrons at filling factor
$\nu = n/(2pn+1)$ and thus the ground state wavefunction \cite{Jain89,Dev92} at these filling factors are
\begin{equation}
\Psi_{\nu}(\Omega_1,\cdots , \Omega_N) = P_{LLL} J \Phi_n(\Omega_1,\cdots , \Omega_N)
\end{equation}
where $\Phi_n$ is the wavefunction for noninteracting CFs in $n$ completely filled lowest $\Lambda$ levels, the Jastrow factor  
\begin{equation}
 J=\prod_{i<j}^N (u_i v_j -v_i u_j)^{2p} ;
\end{equation}
 representing $2p$ number of flux attached to each CF, and $P_{LLL}$ represents the projection into the LLL.

\subsubsection{Lowest Landau Level projection}

Following Jain and Kamilla's projection \cite{JK} into the LLL, the ground state wavefunction can be written as
\begin{equation}
 \Psi_\nu(\Omega_1,\cdots,\Omega_N) = J \Phi_n^{{\rm proj}} (\Omega_1,\cdots, \Omega_N)
\end{equation}
where $\Phi_n^{{\rm proj}}$ is the noninteracting wave function for $n$ completely filled $\Lambda$ levels with projected spherical
harmonic basis states
\begin{eqnarray}
&&Y^{{\rm proj}}_{Q,l,m}(\Omega)= N'_{Qlm}  u^{Q+m} v^{Q-m} \\ \nonumber
&\times&\sum_{s=0}^{l}(-1)^s {{l \choose s}} {{ 2Q+l \choose Q+l-m-s}} P(s,l-s)
 \;\;
\label{single_part_proj}
\end{eqnarray}
where $N'_{Qlm}$ is a normalization constant and $P(s,t)$ for $j$-th particle is defined as
\begin{equation}
 \left( \frac{\partial}{\partial u_j}\right)^s   \left( \frac{\partial}{\partial v_j}\right)^t J_j^p = J_j^p P_j(s,t) 
\end{equation}
with $J = \prod_j J_j^p$ and $P_j[s,t] = \left[ \overline{U}_j^s \overline{V}_j^t 1 \right]$, where
\begin{eqnarray}
\overline{U}_j &=& J_j^{-p} \frac{\partial}{\partial u_j} 
J_j^{p}=p \sum_{k}^{'}\frac{ v_k}{
u_j v_k - v_j u_k} + \frac{\partial}{\partial u_j}, \nonumber \\
\overline{V}_j &=& J_j^{-p} \frac{\partial}{\partial v_j} 
J_j^{p}=p\sum_{k}^{'}\frac{
-  u_k}{u_j v_k - v_j u_k} + \frac{\partial}{\partial v_j}\;.
\end{eqnarray}
which can be simplified as 
\begin{eqnarray}
\overline{U}_j =  f_j(1, 0)+\frac{\partial}{\partial u_j} \\ \nonumber
\overline{V}_j =  f_j(0, 1) +\frac{\partial}{\partial v_j}
\end{eqnarray}
with
\begin{equation}
 f_j(\gamma, \delta) = p\sum_{k}' \left( \frac{v_k}{u_j v_k - u_k v_j} \right)^\gamma
 \left( \frac{-u_k}{u_j v_k - u_k v_j} \right)^\delta.
\end{equation}
where $\sum_k'$ represents $k=j$ is excluded from the sum over $k$.
The derivatives of $f_j(\gamma ,\delta)$ are given by
\begin{eqnarray}
\frac{\partial}{\partial u_j} f_j(\gamma, \delta) &=& -(\gamma + \delta) f_j(\gamma+1, \delta) \\ \nonumber
\frac{\partial}{\partial v_j} f_j(\gamma, \delta) &=& -(\gamma + \delta) f_j(\gamma, \delta+1)
\end{eqnarray}
 Using these derivatives one can determine \cite{JK}
 $P_j(s, t)$ in terms of $f_j(\gamma, \delta) $ and hence
 the wave function $\Psi_\nu$.

In each step  of the Monte Carlo that we use below, we need to calculate $P_j(s,t)$ which requires huge computer
time, specially if we consider large $N$ and $n$.
We find the following recursion relation for $P_j(s,t)$ which has been useful to 
significantly reduce the computing time:
\begin{eqnarray}
   P_j(m,n) &=& \sum_{r=1}^{n+1} (-1)^{r+1} \sum_{s=0}^{m-1} (-1)^{s}  \left (\begin{array}{cc} m-1 \\ s \end{array} \right)
\left [ \sum_{k=0}^{s} (-1)^{k} \right. \nonumber \\
&\times& \left.
 \left (\begin{array}{cc} s \\ k \end{array} \right)   
 \frac{(n+s-k)!}{(n-r+1-k)!} \right] 
 f_j(s+1,r-1) \nonumber \\
 &\times& P_j(m-s-1, n-r+1) 
\end{eqnarray}
 for $m\neq 0$ and
\begin{equation}
 P_j(0,n) = \sum_{r=1}^n (-1)^{r+1} \frac{(n-1)!}{(n-r)!} f_j(0,r) P_j(0, n-r)
\end{equation}
with $P_j(0,0) = 1$.

\subsection{Excited state wave functions}

\subsubsection{Spinless CF-exciton}

  When a CF is excited from a filled $\Lambda$L to an empty $\Lambda$L without changing its spin, a spinless CF-exciton is formed. The composite
fermion wave function of such an excited state at a definite angular momentum $L$ and projection $M=0$ (without losing generality) can be written as
  \begin{eqnarray}
   \Psi_{L,(l,\lambda)}^{(0)} (\Omega_1,\cdots,\Omega_N) &=& J \sum_{m_h} \left( \begin{array}{ccc}
                                                                l+q \,& l+q+\lambda\, & L \\
                                                                 -m_h  & m_h & 0
                                                               \end{array} \right) \nonumber \\
  & \times & \Phi_{n, l,m_h}^{{\rm proj},\lambda} (\Omega_1,\cdots,\Omega_N)
  \end{eqnarray}
where $\Phi_{n, l,m_h}^{{\rm proj},\lambda} (\Omega_1,\cdots,\Omega_N)$ denotes $N$ particle Slater determinant wave function 
with projected spherical harmonic basis states (\ref{single_part_proj}) when all the orbitals except $m_h$ orbital in $l$-th $\Lambda$L amongst the $n$ filled
$\Lambda$Ls for filling factor $\nu=n/(2n+1)$, and $m_h$ orbital in $(l+\lambda)$-th $\Lambda$L are filled by CFs. Here one
CF is excited from $(l,\uparrow) \rightarrow (l+\lambda, \uparrow)$ $\Lambda$ level with $\lambda \geq n-l$ (see Fig.~1),
where $(l,s)$ represents the spin-label $s = \uparrow$ or $\downarrow$ of the $l$-th $\Lambda$L. 

\begin{widetext}

\begin{figure}[h]
\centering
\includegraphics[width=15cm]{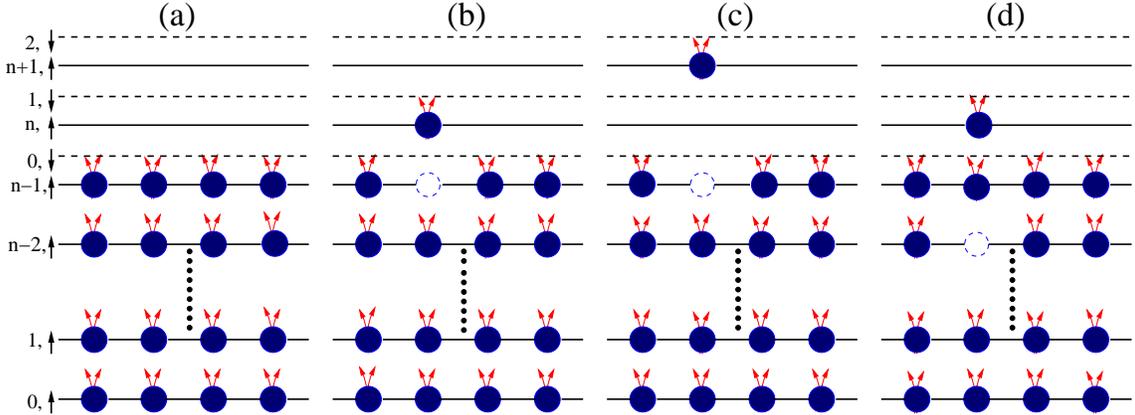}
\caption{(color online) A schematic diagram of the fully polarized FQHE ground state and some of the excited states with 
a spin-zero exciton of CFs. Solid(dashed) lines representing up(down)-spin $\Lambda$Ls are labeled on the left. Filled circles with two
  arrows represent CFs with two vortices attached to each of them; an open circle symbolizes the absence of a CF, {\it i.e.},
  a CF-hole. (a) The spin-polarized ground state of a FQHE filling factor $\nu = n/(2n+1)$ corresponds to  
  completely filled $0,1,\cdots,n-2,n-1$ up-spin $\Lambda$Ls and empty all the down-spin $\Lambda$Ls.
  (b) The level-1$^+$ CF-exciton in which a CF-particle is in the $(n,\uparrow)$ level and a CF-hole is in
  the $(n-1,\uparrow)$ $\Lambda$L. (c) An excitonic state with a level-2$^+$ CF-exciton in which a CF-particle is in the $(n+1,\uparrow)$
  level and a CF-hole is in the $(n-1,\uparrow)$ level. (d) The second level-2$^+$ CF-exciton in which a CF-particle is in the $(n,\uparrow)$ level
  and a CF-hole is in the $(n-2,\uparrow)$ level.}
\label{fig.spin0}
\end{figure}

\begin{figure}[h]
\centering
\includegraphics[width=15cm]{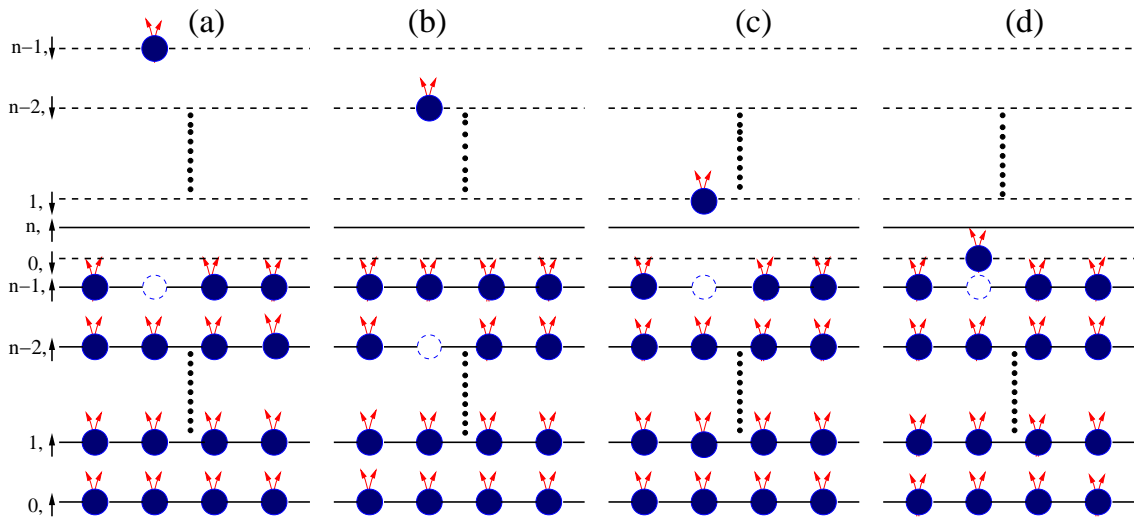}
\caption{(color online) A schematic diagram of some of the spin-one excitons of CFs in the fully polarized ground state of the filling factor $\nu = n/(2n+1)$. 
Filled circles with two
  arrows represent CFs with two vortices attached to each of them; an open circle symbolizes the absence of a CF, {\it i.e.},
  a CF-hole. 
  (a) A level-0 exciton with a CF-particle in the $(n-1,\downarrow)$ level and a CF-hole in the $(n-1,\uparrow)$ level. (b) Another level-0
  exciton with a CF-particle in the $(n-2,\downarrow)$ level and a CF-hole in the $(n-2,\uparrow)$ level. (c) A level-$(n-2)^-$ exciton with
  a CF-particle in the $(1,\downarrow)$ level and CF-hole in the $(n-1,\uparrow)$ level. (d) A level-$(n-1)^-$ exciton with a CF-particle
  in the $(0,\downarrow)$ level and a CF-hole in the $(n-1,\uparrow)$ level.}
\label{fig.spin1}
\end{figure}

\end{widetext}

\subsubsection{Spin-one CF-exciton}

 The spin-one CF-exciton is formed when a CF is excited from a filled $\Lambda$L to an empty $\Lambda$L by reversing its spin.
The corresponding composite fermion wave functions of such excitations for a definite $L$ with projection $M=0$ are given by 
 \begin{eqnarray}
  & & \Psi_{L,(l,\lambda)}^{(1)} (\Omega_1,\cdots,\Omega_N) = J \sum_{m_h} \left( \begin{array}{ccc}
                                                                l+q \,& l+q+\lambda\, & L \\
                                                                 -m_h  & m_h & 0
                                                               \end{array} \right) \nonumber \\
  & \times & \Phi_{n,l,m_h}^{{\rm proj},\lambda} (\Omega_1,\cdots,\Omega_{N-1})Y_{q,l+\lambda,m_h}^{{\rm proj}}(\Omega_N)
  \end{eqnarray}
where $\Phi_{n,l,m_h}^{{\rm proj},\lambda} (\Omega_1,\cdots,\Omega_{N-1})$ denotes $(N-1)$ particle Slater determinant wave function
with projected spherical harmonic basis states (\ref{single_part_proj}) when all the orbitals except $m_h$ orbital in $l$-th $\Lambda$L amongst $n$ filled
$\Lambda$Ls for fiiling factor $\nu = n/(2n+1)$ are filled by CFs. Here one CF is excited from $(l,\uparrow) \rightarrow (l+\lambda,
\downarrow)$ $\Lambda$L (see Fig.2) with $l < n$ and $\lambda \geq -l$. Different levels of excitons are denoted as level-$|\lambda|^{{\rm sign}(\lambda)}$.

\section{Collective Modes}

The composite fermion wave functions for the excited states in a given orbital angular momentum L and spin of CF-excitons for different $(l,\lambda)$
which we collectively label as $\alpha$, are the bare excitonic wave functions $\Psi_{L,\alpha}^{(s)}$ and are not orthogonal, in general. 
All the spin-zero excitonic states get annihilated \cite{Wu95} at $L=1$ upon projection into the lowest LL. There are exactly one spin-zero excitonic state
at $L=2$ and 3 for $\nu=1/3$ shown \cite{Wu95} for small number of particles.
We calculate scalar products $S_{L,\alpha\beta}^{(s)}=\langle\Psi_{L,\alpha}^{(s)}\vert \Psi_{L,\beta}^{(s)}\rangle$, Coulomb matrix elements $\bar{V}^{(s)}_{L,\alpha\beta} = \langle 
\Psi_{L,\alpha}^{(s)}\vert V_C \vert \Psi_{L,\beta}^{(s)}\rangle$, and the ground state energy $E_g = \langle \Psi_\nu \vert V_C \vert
\Psi_\nu \rangle$ by the Monte Carlo method, where $V_C=(e^2/\epsilon)\sum_{i<j}\frac{1}{\vert r_i-r_j\vert}$ is the Coulomb 
interaction with $\epsilon$ and $\vert r_i - r_j\vert$ being the dielectric 
constant and the spherical-chord distance between two particles. The energy gaps of these bare excitons are given by $\bar{\Delta}_{L,\alpha}^{(s)} = \bar{V}_{L,\alpha\alpha}^{(s)}-E_g$.  
We next perform Graham-Schmidt orthogonalization among the states with different $\alpha$ in a 
given $L$ and spin sector. These orthogonal states are labeled as $\Phi_{L,\alpha}^{(s)}$ which are the linear combinations of $\Psi_{L,\beta}^{(s)}$.
We thus obtain Coulomb matrix elements in the orthogonal basis \cite{MJ02}: $V^{(s)}_{L,\alpha\beta} = \langle 
\Phi_{L,\alpha}^{(s)}\vert V(r) \vert \Phi_{L,\beta}^{(s)}\rangle$ which should be obtained using $\bar{V}_{L,\alpha\beta}^{(s)}$ 
and $S_{L,\alpha\beta}^{(s)}$. Finally, we diagonalize the Coulomb matrix $V^{(s)}_{L}$ in this restricted Hilbert space and obtain energy of the 
excited states $E^{(s)}_{L,\alpha}$ and thus the gap for the neutral excitations, $\Delta^{(s)}_{L,\alpha} = E^{(s)}_{L,\alpha}-E_g$. These diagonalized
eigenstates are the linear combination of the excitonic states $\Psi_{L,\alpha}^{(s)}$. For some of the low-lying modes, as we shall see below,
$\bar{\Delta}_{L,\alpha}^{(s)}$ differ substantially from $\Delta_{L,\alpha}^{(s)}$ wherein a large amount of mixing occurs between bare excitonic states.
The linear momentum of the neural collective modes are calculated as $q=L/R$.
Some of the critical energies such as rotons, maxons, long-wavelength and high-momentum modes that are likely to be observed or have been observed are
denoted as $\Delta^s_{k,n}$ in general, where $s=0(1)$ represents spin-zero(one) mode; $k=0,\,\infty,\,Rj,$ and $Mj$ represent long-wavelength, high-momentum,
$j$-th roton, and $j$-th maxon respectively; $n=1,\,2,\,3$ represent the lowest and next higher modes respectively.

\subsection{Filling factor 1/3}
 
\begin{figure}
\vspace{0.5cm}\hspace{1.0cm}
\includegraphics[angle=270,width=8.0cm]{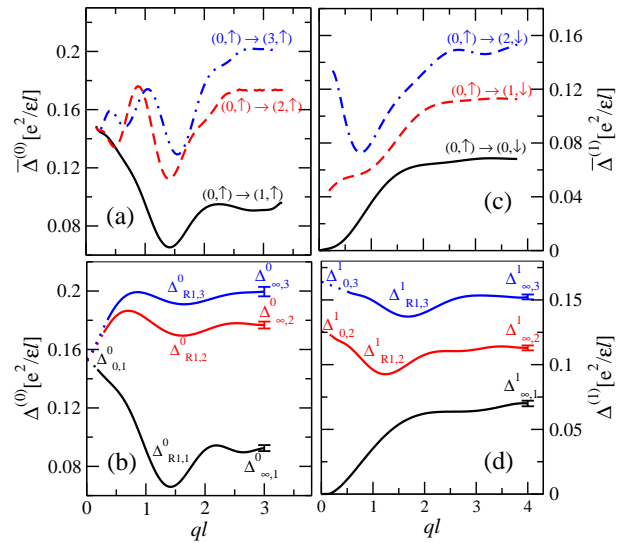}
\caption{(color online) (a) Dispersion of the excitations due to level-1$^+$, level-2$^+$, and level-3$^+$ spin-zero excitons in the fully polarized ground state of filling factor $1/3$
for $N=200$.
(b) The lowest three spin-zero modes determined by considering the mixing of excitons up to level-5. All the modes are extrapolated (dotted lines) up to $q=0$ and they
merge at $q=0$ corresponding to the excitation energy $\Delta^0_0$.
A typical Montecarlo uncertainty for obtaining these modes are shown at the end of these modes. 
Each of these modes exhibit one magneto-roton denoted as $\Delta^0_{R1,1}$, $\Delta^0_{R1,2}$, and $\Delta^0_{R1,3}$. Their energies at the high momentum
limit are denoted as $\Delta^0_{\infty,1}$, $\Delta^0_{\infty,2}$, and $\Delta^0_{\infty,3}$.
(c) Three lowest spin-one excitonic modes, {\it viz}., $(0,\uparrow) \to (0,\downarrow)$, $(0,\uparrow) \to (1,\downarrow)$, and $(0,\uparrow) \to (2,\downarrow)$ 
for $N=100$.
(d) Three lowest spin-one modes arising from the mixing of spin-one excitonic modes up to level-4$^+$. The lowest mode is the spin-wave mode. Each of the other 
two spin-flip modes exhibit one spin-roton denoted as $\Delta^1_{R1,2}$ and $\Delta^1_{R1,3}$. The extrapolated energies of these two spin-flip modes up to $q=0$
are denoted as $\Delta^1_{0,2}$ and $\Delta^1_{0,3}$. The high-momentum limit of the energies of these three modes are denoted as $\Delta^1_{\infty,1}$,
$\Delta^1_{\infty,2}$, and $\Delta^1_{\infty,3}$. The total energy for these spin-one modes are $\omega = E_z + \Delta^{(1)}$. }
\label{Fig_spec_13}
\end{figure}

In the case of spin-zero excitations at filling factor $\nu =1/3$, the following characteristics which are independent of $N$ are noteworthy. 
(i) There is no excited state 
at $L=1$ as the state $\Psi^{(0)}_{1,(0,1)}$ is identically zero. (ii) Only one linearly independent excited state exists both
at L=2 and L=3 since $\langle \Psi^{(0)}_{2,(0,1)} \vert \Psi^{(0)}_{2,(0,2)} \rangle = 1$ and $\langle \Psi^{(0)}_{3,(0,1)} \vert \Psi^{(0)}_{3,(0,2)}
 \rangle = \langle \Psi^{(0)}_{3,(0,1)} \vert \Psi^{(0)}_{3,(0,3)}\rangle =1$. (iii) Although there are four states available at $L=4$, {\em viz.},
$\Psi^{(0)}_{4,(0,1)}$, $\Psi^{(0)}_{4,(0,2)}$, $\Psi^{(0)}_{4,(0,3)}$, and $\Psi^{(0)}_{4,(0,4)}$, only two linearly independent states exist. 
(iv) The number of linearly independent states increases with $L$ and becomes same with the number of possible excitonic states at large $L$.
Figure \ref{Fig_spec_13}(a)shows dispersions of the lowest three excitonic modes, {\it viz.}, $(0,\uparrow) \to (1,\uparrow)$,
$(0,\uparrow) \to (2,\uparrow)$, and $(0,\uparrow) \to (3,\uparrow)$ whose energies have been calculated using excitonic wave functions
$\Psi^{(0)}_{L,(0,1)}$, $\Psi^{(0)}_{L,(0,2)}$, $\Psi^{(0)}_{L,(0,3)}$. The lowest mode, {\it i.e.}, level-1$^+$ excitonic mode
has been reported in earlier studies \cite{Dev92,JK,Scarola00}. It is now 
interesting to see if the other higher excitonic modes influence the lowest mode or vice versa, especially in the low momentum region where
the energies of all these three excitonic modes are close.
We explicitly calculate $\Delta_{L,\alpha}^{(0)}$ for the lowest five modes ($\alpha$=1--5) considering up to level-5$^+$
excitons of CFs and have shown the lowest three modes in Fig.~\ref{Fig_spec_13}(b). 
(Other two higher energy modes could not be distinguished due to uncertainty arising from Monte Carlo evaluation of the energies.)
We calculate these modes for $L\geq 5$ and up to 200 particles, and notice that energies of the two higher energy modes decrease from their respective
maxima while energy of the lowest mode keeps on increasing from its minimum on lowering $L$. 
We then extrapolate these modes up to $q=0$ by exploiting the property that only one mode exists near $q=0$ since only one
linearly independent excitonic state exists at $L=2$ and 3, no matter what the value of $N$ is. The energy corresponding to $q=0$ mode
is denoted as $\Delta^0_{0,1}$.
This explains the observed mode-splitting \cite{Hirjibehedin05} at $\nu=1/3$ in an ILS. In a recent paper, Yang and Haldane \cite{Haldane14}
proposed that the observed modes \cite{Majumder11_1} for different levels of CF excitons may be thought as different orbits of a quasihole of charge $-e/3$
orbiting around a quasiparticle of charge $2e/3$ so that the composites describe a family of $e/3$ quasiparticle states. A roton minimum has been
developed in each of the three collective modes shown in Fig.~\ref{Fig_spec_13}(b); the corresponding energies are denoted respectively as $\Delta^0_{R1,1}$,
$\Delta^0_{R1,2}$, and $\Delta^0_{R1,3}$. The energies corresponding to the high-momentum limit of these modes are denoted as $\Delta^0_{\infty,1}$,
$\Delta^0_{\infty,2}$, and $\Delta^0_{\infty,3}$.

Figure \ref{Fig_spec_13}(c) shows spin-one excitation modes calculated by considering spin-1 excitonic wave functions $\Psi^{(1)}_{L,(0,0)}$, $\Psi^{(1)}_{L,(0,1)}$, and
$\Psi^{(1)}_{L,(0,2)}$ corresponding to the respective level-0, level-1$^+$, and level-2$^+$ excitons 
$(0,\uparrow)\to (0,\downarrow)$, $(0,\uparrow) \to (1,\downarrow)$, and $(0,\uparrow) \to (2,\downarrow)$
for all $L$, excepting $L=0$ where only the first state exists and $L=1$ where the first two states are possible only.
The level-0  mode is a conventional spin-wave for a ferromagnetic ground state; the level-1$^+$ mode does not have any well-formed ``spin-roton'' minimum;
the level-2$^+$ mode has a spin-roton minimum. 
We consider up to level-5$^+$ excitons to calculate $\Delta^{(1)}_{L,\alpha}$ and the lowest three modes are shown in Fig.~\ref{Fig_spec_13}(d).
The lowest mode is essentially the level-0 excitonic mode as it does not mix with the other higher levels of excitons.
The other excitonic modes mix, especially at the small momentum region and we find one well-formed spin-roton each in both of the next higher energy modes.
The respective spin-roton energies are denoted as $\Delta^1_{R1,2}$ and $\Delta^1_{R1,3}$ which have been observed \cite{Majumder11_1}. The respective
energies of all the three modes at the high-momentum limit are denoted as $\Delta^1_{\infty,1}$, $\Delta^1_{\infty,2}$, and $\Delta^1_{\infty,3}$.
Extrapolated energy up to $q=0$ for the two higher energy modes are denoted as $\Delta^1_{0,2}$ and $\Delta^1_{0,3}$.

\subsection{Filling factor 2/5}

\begin{figure}
\vspace{0.5cm}\hspace{1.0cm}
\includegraphics[angle=270,width=8.5cm]{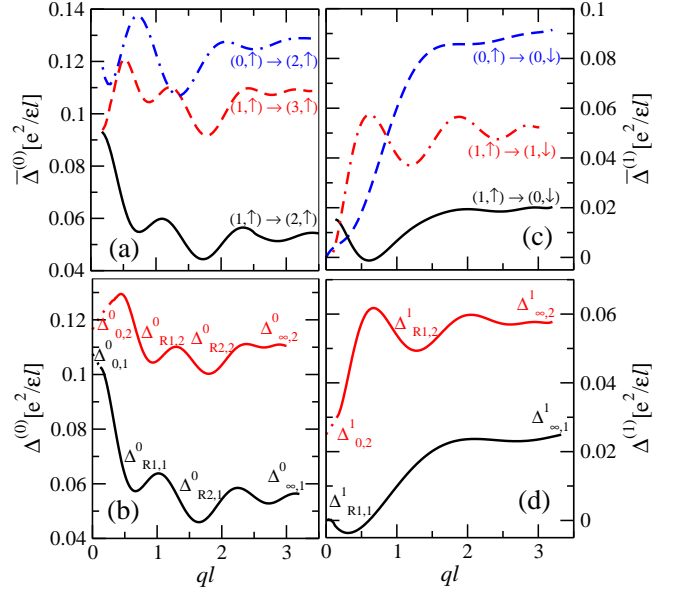}
\caption{(color online) Spin-zero and spin-one modes at the fully polarized filling factor $2/5$. (a) The level-1$^+$ and level-2$^+$ spin-zero excitonic modes for $N=200$.
(b) Two lowest spin-zero modes arising from the mixing up to level-$3^+$ excitons. Both the modes exhibit two rotons and they are denoted as $\Delta^0_{R1,1}$,
$\Delta^0_{R2,1}$, $\Delta^0_{R2,1}$ and $\Delta^0_{R2,2}$. The dotted lines denote the extrapolation of these modes up to $q=0$ where their energies are denoted as
$\Delta^0_{0,1}$ and $\Delta^0_{0,2}$. The energies of these modes at high momentum limit are denoted as $\Delta^0_{\infty,1}$ and $\Delta^0_{\infty,2}$.
(c) Two level-0 and one level-$1^-$ spin-one excitonic modes for $N=100$. The former two behave as spin waves at long wavelength, and the latter one displays a spin-roton.
(d) Two lowest spin-one modes obtained by mixing up to level-1$^+$ spin-one excitonic modes. The higher mode has one spin-roton denoted as $\Delta^1_{R1,2}$.
The energies of these modes at $q\to 0$ are zero and $\Delta^1_{0,2}$ respectively; high-momentum limit of these modes are denoted as
$\Delta^1_{\infty,1}$ and $\Delta^1_{\infty,2}$. The total energy for spin-one excitations are given by $\omega = E_z +\Delta^{(1)}$.}
\label{Fig_spec_25}
\end{figure}

In the fully polarized ground state of 2/5, $(0,\uparrow)$ and $(1,\uparrow)$ $\Lambda$Ls are filled. We consider
up to level-4$^+$ spin-zero CF-excitons that amounts to seven excitons ($\alpha=$1--7) in all. The three bare modes corresponding to level-1$^+$ and level-2$^+$ excitons
$(1,\uparrow)\rightarrow (2,\uparrow)$, $(1,\uparrow) \rightarrow (3,\uparrow)$, and $(0,\uparrow)\rightarrow (2,\uparrow)$
whose respective energies are calculated using wave functions
 $\Psi^{(0)}_{L,(1,1)}$, $\Psi^{(0)}_{L,(1,2)}$, and $\Psi^{(0)}_{L,(0,2)}$ are shown in Fig.~\ref{Fig_spec_25}(a). 
 Unlike 1/3 state, the excitonic wave functions here are not identical at $L=2$ and $L=3$. This rules out the merging of the actual modes in the thermodynamic limit. 
Figure \ref{Fig_spec_25}(b) depicts the lowest two modes determined by mixing all the seven excitons. Two roton minima are developed in both the modes and
these are denoted as $\Delta^0_{R1,1},\, \Delta^0_{R2,1},\, \Delta^0_{R1,2}$ and $\Delta^0_{R2,2}$; both the modes
come closer in the long-wavelength limit but their energy separation remains finite as $\Delta^0_{0,1}\neq \Delta^0_{0,2}$; the energies of these modes at large momenta are
shown as $\Delta^0_{\infty,1}$ and $\Delta^0_{\infty,2}$ respectively.

In spin-one excitations, we consider two level-0 excitons, $(0,\uparrow)\to (0,\downarrow)$ and $(1,\uparrow) \to (1\downarrow)$, one level-1$^-$ exciton 
$(1,\uparrow)\to (0,\downarrow)$, and two level-1$^+$ exciton $(1,\uparrow) \to (2,\downarrow)$ and $(0,\uparrow) \to (1,\downarrow)$ and two level-2$^+$ 
excitons $(1,\uparrow) \to (3,\downarrow)$ and $(0,\uparrow) \to (2,\downarrow)$, {\it i.e.}, $\alpha=$1--7. 
The wave functions these respective excitons are $\Psi^{(1)}_{L,(0,0)}$, $\Psi^{(1)}_{L,(1,0)}$, $\Psi^{(1)}_{L,(1,-1)}$, $\Psi^{(1)}_{L,(1,1)}$,
$\Psi^{(1)}_{L,(0,1)}$, $\Psi^{(1)}_{L,(1,2)}$, and $\Psi^{(1)}_{L,(0,2)}$.
The level-0 excitons correspond to spin-waves,
{\em a la}, itinerant ferromagnetic systems, {\it i.e.}, no Coulomb energy needed for flipping a spin in infinite wavelength limit. The level-1$^-$ excitons describes 
spin-flip excitations with lowering $\Lambda$L and hence the excitation (Coulomb) energy is negative for a region of momentum and thereby formation of a
spin-roton minimum. These three modes are shown in Fig.~\ref{Fig_spec_25}(c). However, the mixing between these three modes generates a novel spin-wave mode
in which long-wavelength part is dominated by level-0 excitons and high-momentum regions are predominantly level-1$^-$ exciton, and the intermediate region
is a mixture of level-0 and level-$1^-$ excitons. Therefore the lowest spin-one mode begins with zero Coulomb energy at zero momentum and then gradually lowering
its energy until forming a spin-roton which we label as $\Delta^1_{R1,1}$ in Fig.~\ref{Fig_spec_25}(d), followed by gradually increase to a positive energy at large
momentum region denoted as $\Delta^1_{\infty,1}$. The next higher mode that we have shown in Fig.~\ref{Fig_spec_25}(d) is predominantly the level-0, {\it viz}, 
$(1,\uparrow) \to (1\downarrow)$ exciton at higher momentum and mixing of all the level-0 and level-1$^-$ excitons occurs at lower momenta. This mode also shows a roton minimum denoted as
$\Delta^1_{R1,2}$, thermodynamically extended energy $\Delta^1_{0,2}$ at long wave-length, and flatness at large momentum with energy $\Delta^1_{\infty,2}$.
There is no significant mixing of other higher level excitons occurs for these two lowest spin-one modes.

\subsection{Filling factor 3/7}

\begin{figure}[h]
\vspace{0.5cm}\hspace{0.5cm}
\includegraphics[angle=270,width=8.5cm]{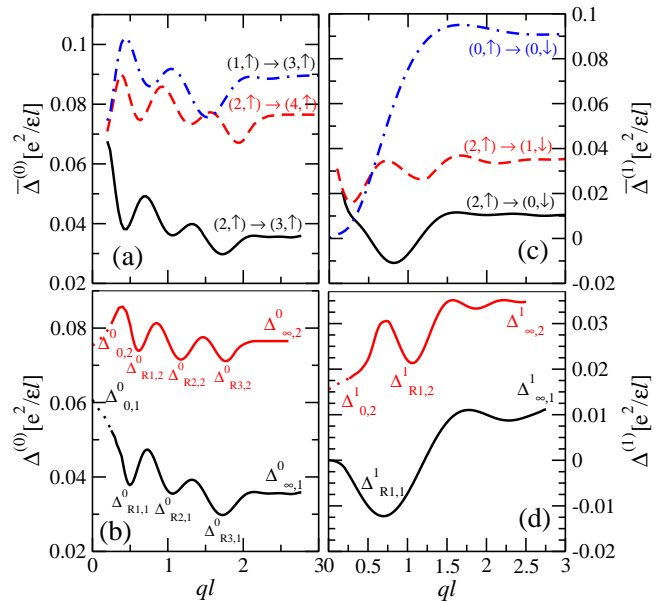}
\caption{(color online) Spin-zero and spin-one modes at the fully polarized filling factor $3/7$. (a) The level-1$^+$ and level-2$^+$ spin-zero excitonic modes for $N=136$.
(b) Two lowest spin-zero modes arising from the mixing up to level-$3^+$ excitons. Both the modes exhibit three rotons and they are denoted as $\Delta^0_{R1,1}$,
$\Delta^0_{R2,1}$, $\Delta^0_{R3,1}$, $\Delta^0_{R2,1}$, $\Delta^0_{R3,1}$ and $\Delta^0_{R3,2}$. The dotted lines denote the extrapolation of these modes up to $q=0$ where their energies are denoted as
$\Delta^0_{0,1}$ and $\Delta^0_{0,2}$. The energies of these modes at high momentum limit are denoted as $\Delta^0_{\infty,1}$ and $\Delta^0_{\infty,2}$.
(c) One level-0, one level-$1^-$, and one level-$2^-$ spin-one excitonic modes for $N=136$. 
(d) Two lowest spin-one modes obtained by mixing up to level-1$^+$ spin-one excitonic modes. Both the modes have one spin-roton denoted as $\Delta^1_{R1,1}$
$\Delta^1_{R1,2}$. The energies of these modes at $q\to 0$ are zero and $\Delta^1_{0,2}$ respectively; high-momentum limit of these modes are denoted as
$\Delta^1_{\infty,1}$ and $\Delta^1_{\infty,2}$. The total energy for spin-one excitations are given by $\omega = E_z +\Delta^{(1)}$. }
\label{Fig_spec_37}
\end{figure}

In the fully polarized ground state of filling factor $3/7$, $(0,\uparrow)$, $(1,\uparrow)$, and $(2,\uparrow)$ $\Lambda$L's are 
completely filled. We consider six excitonic wavefunctions $(\alpha =1$--6), {\em viz}, $\Psi^{(0)}_{L,(2,1)}$,
$\Psi^{(0)}_{L,(2,2)}$, $\Psi^{(0)}_{L,(1,2)}$, $\Psi^{(0)}_{L,(2,3)}$,
$\Psi^{(0)}_{L,(1,3)}$, and $\Psi^{(0)}_{L,(0,3)}$ corresponding to respective 
one level-1$^+$ exciton  $(2,\uparrow)\rightarrow (3,\uparrow)$, two level-2$^+$ excitons $(2,\uparrow)
\rightarrow (4,\uparrow)$ and $(1,\uparrow) \rightarrow (3,\uparrow)$, and three level-$3^+$ excitons $(2,\uparrow)
\rightarrow (5,\uparrow)$, $(1,\uparrow)  \rightarrow (4,\uparrow)$, and
$(0,\uparrow) \rightarrow (3,\uparrow)$. The dispersion due to level-1$^+$ excitons is available in literature \cite{JK}. Figure \ref{Fig_spec_37}(a) shows 
the modes for level-1$^+$ and level-2$^+$ excitons. The mixing of these modes and the modes due to other higher level excitons that we have considered 
causes renormalization of the low-lying modes. The lowest two modes are shown in Fig.~\ref{Fig_spec_37}(b). Each of these modes have three roton minima that
are denoted as $\Delta^0_{R1,1}$, $\Delta^0_{R2,1}$, $\Delta^0_{R3,1}$, $\Delta^0_{R1,2}$, $\Delta^0_{R2,2}$, and $\Delta^0_{R3,2}$. The finite energy 
separation between these two modes at the long wave-length limit (shown as extrapolation of the dispersion) is visible; the respective energies are 
represented by $\Delta^0_{0,1}$ and $\Delta^0_{0,2}$. The high momentum limit of the energies of these modes are represented by $\Delta^0_{\infty,1}$
and $\Delta^0_{\infty ,2}$.

We consider three level-0 excitons $(2,\uparrow) \to (2,\downarrow)$, $(1,\uparrow) \to (1,\downarrow)$, $(0,\uparrow) \to (0,\downarrow)$, one level-$2^-$
exciton $(2,\uparrow) \to (0,\downarrow)$, two level-$1^-$ excitons $(2,\uparrow) \to (1,\downarrow)$ and $(1,\uparrow) \to (0,\downarrow)$, and three
level-$1^+$ excitons $(2,\uparrow) \to (3,\downarrow)$, $(1,\uparrow) \to (2,\downarrow)$, and $(0,\uparrow) \to (1,\downarrow)$ for determining low-lying
spin-one modes in the fully polarized ground state at the filling factor 3/7. 
The wave functions for these respective excitons are $\Psi^{(1)}_{L,(2,0)}$, $\Psi^{(1)}_{L,(1,0)}$, $\Psi^{(1)}_{L,(0,0)}$, $\Psi^{(1)}_{L,(2,-1)}$,
$\Psi^{(1)}_{L,(1,-1)}$, $\Psi^{(1)}_{L,(2,1)}$, $\Psi^{(1)}_{L,(1,1)}$, and $\Psi^{(1)}_{L,(0,1)}$.
The modes for one each of level-0, level-$1^-$, and level-$2^-$ are shown 
in Fig.~\ref{Fig_spec_37}(c). While the level-0 modes suggest the presence of spin-wave mode, {\it i.e.}, $\Delta^{(1)} (q\to 0) =0$, the lowering of $\Lambda$Ls
in level-$2^-$ and level-$1^-$ modes exhibit the formation of spin-roton minima at an energy lower than the Zeemen energy. The mixing of all these modes along 
with the modes of higher level excitons generate low-lying spin-one modes. Two lowest spin-one modes are shown in Fig.~\ref{Fig_spec_37}(d) that we have obtained
considering up to level-$1^+$ excitons. The characteristic of the lowest mode can be represented by three regions in which the low-momentum region is dominated
by the spin-wave modes, {\it i.e.}, level-0 excitons, high-momentum region is dominated by spin-flip mode with maximum possible lowering of $\Lambda$Ls, i.e., level-2$^-$
exciton and the intermediate regime is due to the nontrivial mixture of all the level-0, level-2$^-$, and level-1$^-$ excitons. 
The high-momentum region of the next higher mode is due to the mixing of two level-$1^-$ excitons. Its low and and intermediate regions are the result of the 
mixing of all the modes considered. Both the modes have one spin-roton denoted by $\Delta^1_{R1,1}$ and $\Delta^1_{R1,2}$ of which $\Delta^1_{R1,1}$ here (also
for filling factors 2/5 above and 4/9 below) is particularly
interesting because the excitation energy is less than the Zeeman energy. This suggests the possibility of excitations at sub-Zeeman energies and the fully
spin-polarized ground state at sufficiently small Zeeman energies is unstable.
The energies at the high-momentum limits of these two modes are denoted as $\Delta^1_{\infty, 1}$ and $\Delta^1_{\infty,2}$. The energy of the long-wavelength limit
of the higher mode is denoted as $\Delta^1_{0,2}$.

\subsection{Filling factor 4/9}

\begin{figure}
\vspace{0.5cm}\hspace{0.5cm}
\includegraphics[angle=270,width=8.0cm]{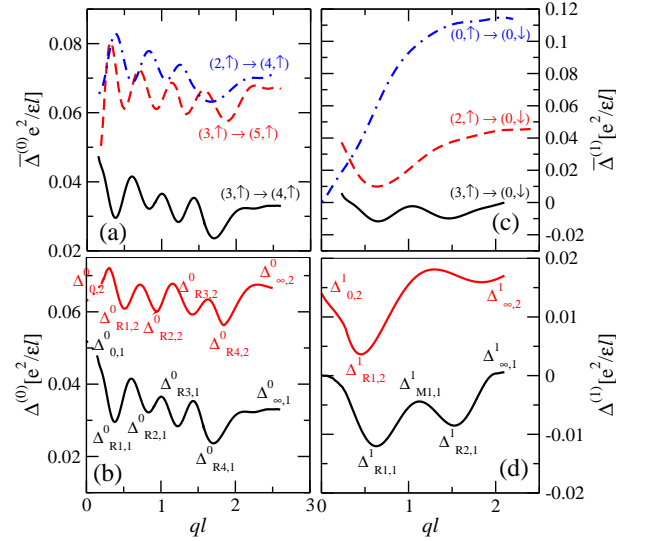}
\caption{(color online) Spin-zero and spin-one modes at the fully polarized filling factor $4/9$. (a) The level-1$^+$ and level-2$^+$ spin-zero excitonic modes for $N=160$.
(b) Two lowest spin-zero modes arising from the mixing up to level-$3^+$ excitons. Both the modes exhibit three rotons and they are denoted as $\Delta^0_{R1,1}$,
$\Delta^0_{R2,1}$, $\Delta^0_{R3,1}$, $\Delta^0_{R4,1}$, $\Delta^0_{R1,2}$, $\Delta^0_{R2,2}$, $\Delta^0_{R3,2}$, and $\Delta^0_{R4,2}$. 
The dotted lines denote the extrapolation of these modes up to $q=0$ where their energies are denoted as
$\Delta^0_{0,1}$ and $\Delta^0_{0,2}$. The energies of these modes at high momentum limit are denoted as $\Delta^0_{\infty,1}$ and $\Delta^0_{\infty,2}$.
(c) One level-0, one level-$2^-$, and one level-$3^-$ spin-one excitonic modes for $N=160$. 
(d) Two lowest spin-one modes obtained by mixing up to level-1$^+$ spin-one excitonic modes. The lowest mode has two spin-rotons denoted as $\Delta^1_{R1,1}$ and
$\Delta^1_{R2,1}$, and the next higher mode has one spin-roton denoted as $\Delta^1_{R1,2}$. The lowest mode has a maxon denoted by $\Delta^1_{M1,1}$ as well.
The energies of these modes at $q\to 0$ are zero and $\Delta^1_{0,2}$ respectively; high-momentum limit of these modes are denoted as
$\Delta^1_{\infty,1}$ and $\Delta^1_{\infty,2}$. The total energy for spin-one excitations are given by $\omega = E_z +\Delta^{(1)}$.}
\label{Fig_spec_49}
\end{figure}

In the fully polarized ground state of $\nu =4/9$, $(0,\uparrow)$, $(1,\uparrow)$, $(2,\uparrow)$, and $(3,\uparrow)$ $\Lambda$Ls are fully filled and all the 
other $\Lambda$Ls are completely empty. We consider wave functions corresponding to six $(\alpha =1$--6) spin-zero excitons. They are $\Psi^{(0)}_{L,(3,1)}$,
$\Psi^{(0)}_{L,(3,2)}$, $\Psi^{(0)}_{L,(2,2)}$, $\Psi^{(0)}_{L,(3,3)}$, $\Psi^0_{L,(2,3)}$, and $\Psi^0_{L,(1,3)}$ corresponding to respective one level-1$^+$
exciton $(3,\uparrow)\to (4,\uparrow)$, two level-$2^+$ excitons $(3,\uparrow) \to (5,\uparrow)$ and $(2,\uparrow) \to (4,\uparrow)$, and three level-$3^+$
excitons $(3,\uparrow) \to (6,\uparrow)$, $(2,\uparrow) \to (5,\uparrow)$, and $(1,\uparrow) \to (4,\uparrow)$. The dispersion for level-$1^+$ exciton has been
studied before. Figure \ref{Fig_spec_49}(a) shows the dispersion for level-$1^+$ and level-$2^+$ excitons. The mixing of these modes and the modes due to level-$3^+$
excitons considered here determines the low-lying spin-zero modes at $\nu=4/9$. The lowest two modes are shown in Fig.~\ref{Fig_spec_49}(b). The lowest mode
does not get renormalized by the mixing and thus it is mostly due to level-$1^+$ exciton. The second mode arises due to the mixing more of level-$2^+$ and
less of level-$3^+$ excitons. These modes have four rotons each that are denoted as 
$\Delta^0_{R1,1},\cdots,\Delta^0_{R4,1}$ and $\Delta^0_{R1,2},\cdots,\Delta^0_{R4,2}$ respectively. The long-wavelength extrapolation of these modes provide
energy separation of these modes and they are denoted as $\Delta^0_{0,1}$ and $\Delta^0_{0,2}$. The high momentum limit of the energies of these modes are represented
as $\Delta^0_{\infty,1}$ and $\Delta^0_{\infty,2}$.

For determining low-lying spin-one modes in the fully polarized phase at $\nu=4/9$, we consider four level-0 excitons: $(3,\uparrow) \to (3,\downarrow)$, $(2,\uparrow) \to (2,\downarrow)$, $(1,\uparrow) \to (1,\downarrow)$, and $(0,\uparrow)
\to (0,\downarrow)$, three level-$1^-$ excitons: $(3,\uparrow) \to (2,\downarrow)$, $(2,\uparrow) \to (1,\downarrow)$, and $(1,\uparrow) \to (0,\downarrow)$,
two level-$2^-$ excitons: $(3,\uparrow) \to (1,\downarrow)$ and $(2,\uparrow) \to (0,\downarrow)$, and one level-$3^-$ exciton: $(3,\uparrow) \to (0,\downarrow)$
whose respective wave functions are $\Psi^{(1)}_{L,(3,0)}$, $\Psi^{(1)}_{L,(2,0)}$, $\Psi^{(1)}_{L,(1,0)}$, $\Psi^{(1)}_{L,(0,0)}$, $\Psi^{(1)}_{L,(3,-1)}$,
$\Psi^{(1)}_{L,(2,-1)}$, $\Psi^{(1)}_{L,(1,-1)}$, $\Psi^{(1)}_{L,(3,-2)}$, $\Psi^{(1)}_{L,(2,-2)}$, and $\Psi^{(1)}_{L,(3,-3)}$.
The dispersions corresponding to $(0,\uparrow) \to (0,\downarrow)$, $(2,\uparrow) \to (0,\downarrow)$ and 
$(3,\uparrow) \to (0,\downarrow)$ excitons are shown in Fig.~\ref{Fig_spec_49}(c).
While dispersions of level-0 excitons suggest the presence of spin-wave mode, {\it i.e.}, $\Delta (q\to 0)=0$, the lowering of $\Lambda$Ls for level-$3^-$ and
level-$2^-$ excitons correspond to the formation of spin-rotons. We obtain two lowest spin-one modes (Fig.~\ref{Fig_spec_49}(d)) by mixing all the above excitonic modes.
As in the case of $\nu=3/7$, the lowest spin-one mode here can also be characterized by three regions in which the low momentum region is dominated by spin-wave mode,
{\it i.e.}, level-0 excitons, high momentum region is dominated by spin-flip mode with maximum lowering of $\Lambda$Ls, {\it i.e.}, level-$3^-$ excitons, and the intermediate
regime is the nontrivial mixing of the level-0 excitons and all the excitons with excitaions by lowering $\Lambda$Ls.
The lowest mode has two spin-rotons denoted as $\Delta^1_{R1,1}$ and $\Delta^1_{R2,1}$ and a maxon $\Delta^1_{M1,1}$, and the next higher mode has one spin-roton 
$\Delta^1_{R1,2}$. The energies of these two modes at high-momentum limits are denoted as $\Delta^1_{\infty ,1}$ and $\Delta^1_{\infty, 2}$. 
The energy $\Delta^1_{\infty ,1}$ here and also for $\nu =1/3,\, 2/5$ and 3/7 are the corresponding energies of the spin-reversed gap for charged 
excitations \cite{Mandal02}.
The energy of the long-wavelength limit of the higher mode is denoted as $\Delta^1_{0,2}$.

\section{Effect of Finite Width of Quantum Wells}

\subsection{Effective Interaction Potential}

We here review the method \cite{Ortalano97,Meshkini} of determining effective two-dimensional potential $V_{{\rm eff}}(r)$ due to finite extent
of the single particle wave function along transverse direction in a quantum well of thickness $d$.  This can be
straightforwardly determined as
\begin{equation}
 V_{{\rm eff}} (r) = \int dz_1 \int dz_2 \vert \xi(z_1)\vert^2 \vert \xi(z_2)\vert^2 \frac{e^2}{\epsilon \sqrt{r^2 + (z_1-z_2)^2}}
 \label{Effective_potential}
\end{equation}
where $z_1$ and $z_2$ are the transverse coordinates of two particles, and $\xi (z)$ is the lowest subbband solution of the
Schrodinger equation
\begin{equation}
 \left( -\frac{\hbar^2}{2m}\frac{d^2}{dz^2} + U (z) \right) \xi (z) = E \xi (z)
 \label{Schrodinger}
\end{equation}
where the effective one electron potential energy $U(z)$:
\begin{equation}
 U(z) = U_W(z)+U_H(z)+U_{ex}(z).
\end{equation}
Here $U_W(z)$ is the quantum well confinement potential, $U_H(z)$ is the self-consistent Hartree potential satisfying
Poisson's equation
\begin{equation}
 \frac{d^2U_H(z)}{dz^2} = -\frac{4\pi e^2}{\epsilon}\left[ n(z)-n_I(z) \right]
 \label{Hartree_pot}
\end{equation}
with $n(z)=(m/\pi)\vert \xi(z) \vert^2$ 
being the electron density computed
from the effective single-particle lowest subband wave function, and $n_I(z)$ the density of donar ions,
and $U_{ex}(z)$ is the exchange correlation potential. It is assumed that the back-ground charge density is uniform, i.e.,
$n_I (z) = n_e$, mean electron density in the quantum well. Many-body effects beyond the mean-field Hartree approximation is
considered by using density functional theory in the LDA approximation. We use Hedin and Lundqvist \cite{Hedin71} parametrization of
the exchange potential as
\begin{equation}
 U_{ex}(z) = -[1+0.7734x\ln (1+x^{-1})]\left(\frac{2}{\pi\beta r_s}\right){\cal R},
 \label{Exchange_pot}
\end{equation}
where $\beta = (4/9\pi)^{1/3}$, $x=r_s/21$, and $r_s = (4\pi a^{*3}n(z)/3)^{1/3}$ with 
$a^*$ and ${\cal R}$ being the effective Bohr radius and the effective Rydberg respectively.

\begin{figure}
\vspace{0.5cm}\hspace{0.5cm}
\includegraphics[angle=270,width=8.2cm]{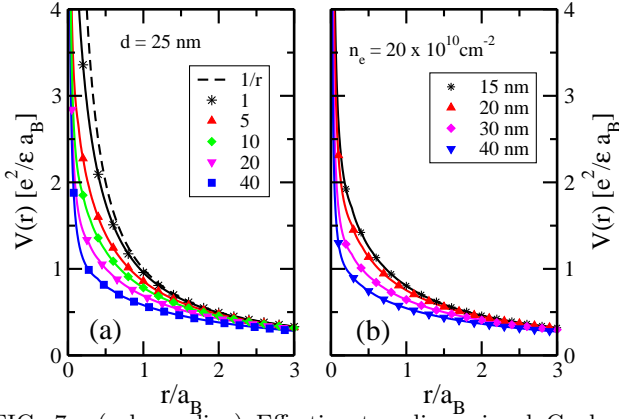}
\vspace{-0.4cm}
\caption{(color online) Effective two-dimensional Coulomb potential $V_{{\rm eff}}(r)$ obtained by local density approximation in quantum wells of finite transverse width.
 Dotted lines for bare Coulomb potential. $V_{{\rm eff}}(r)$  (a) for various electron densities in the unit of $10^{10}\,{\rm cm}^{-2}$ at a fixed
 width $d=25$nm, and (b) for various widths of quantum wells at a fixed electron density $n_e = 2\times 10^{11}\,{\rm cm}^{-2}$. }
\label{fig.LDA_pot}
\end{figure}

The self-consistent evaluation procedure of $\xi (z)$ is then started with an initial guess of $n(z)$,
followed by the evaluation of $U_H(z)$ and $U_{ex}(z)$ using Eqs.~(\ref{Hartree_pot}) and (\ref{Exchange_pot}).  
The Schrodinger equation (\ref{Schrodinger}) is
then used to obtain $\xi (z)$ and hence $n(z)$. The trial value of $n(z)$ is chosen for the next step
as the sum of a chosen fraction of old $n(z)$ and the remaining fraction of new $n(z)$. This procedure is
continued until the parameter which is the ratio of the integrated absolute value of the deviation of
$n(z)$ and the integrated value of old $n(z)$ becomes less than a desired tolerance. Once $\xi (z)$ is
self-consistently determined, $V_{{\rm eff}}$ can readily be evaluated {\em via} Eq.~(\ref{Effective_potential}).

Figure \ref{fig.LDA_pot} shows $V_{{\rm eff}}(r)$ for different values of electron densities in the quantum wells and
their widths. Clearly, the short distance part of the bare Coulomb interaction decreases more on increasing the widths 
as well as electron densities. It, however, will not have any remarkable qualitative influence on the collective modes, although
its quantitative dependence on the energies of the collective modes is prominent as we will see below.

 \subsection{Finite Width Correction to Critical Energies}
 
  In inelastic light scattering experiments, a typical momentum transfer that occur is $q \lesssim 0.1 \ell^{-1}$, which is thus capable of determining energies 
  of long-wavelength neutral collective modes. The presence of impurity in the systems breaks translational invariance and hence
  one expects of finding resonance in the spectra corresponding to the excitation energies at which the density of states is very high \cite{Platzman94}. These
  are the energies for rotons, maxons, and high-momentum limits. Energies of the maxons are typically found to be very close to the energies of high-momentum limit
  or one of the rotons and thus it cannot be distinguished in an ILS experiment. We here thus estimate the finite thickness dependent of these critical energies (neglecting
  maxons except for one case) for all the modes. The observation of some of these critical energies have already been reported \cite{Majumder11_1,Majumder11_2}.
  
\begin{figure}[h]
\vspace{1.0cm}\hspace{0.5cm}
\includegraphics[angle=270,width=8.0cm]{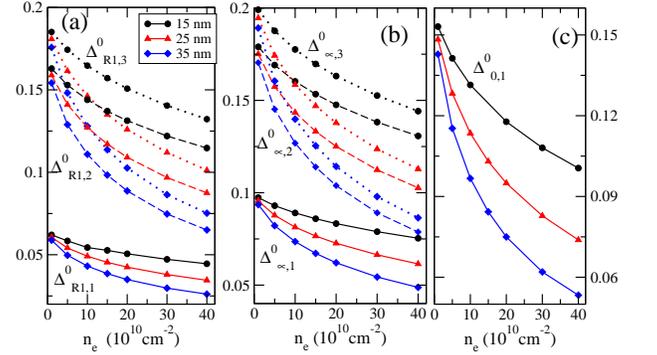}
\caption{(color online) The variation of critical energies corresponding to spin-zero modes (Fig.~\ref{Fig_spec_13}(b)) 
with electron densities and width of the quantum wells at filling factor 1/3.
Critical energies in the unit of $e^2/(\epsilon \ell)$ for (a) rotons: $\Delta^0_{R1,1}$, $\Delta^0_{R1,2}$, and $\Delta^0_{R1,3}$; (b) high-momentum modes: $\Delta^0_{\infty,1}$,
$\Delta^0_{\infty,2}$, and $\Delta^0_{\infty,3}$; (c) long-wavelength mode: $\Delta^0_{0,1}$.}
\label{Fig_cm_13}
\end{figure}

\begin{figure}[ht]
\vspace{0.5cm}\hspace{0.5cm}
\includegraphics[angle=270,width=8.0cm]{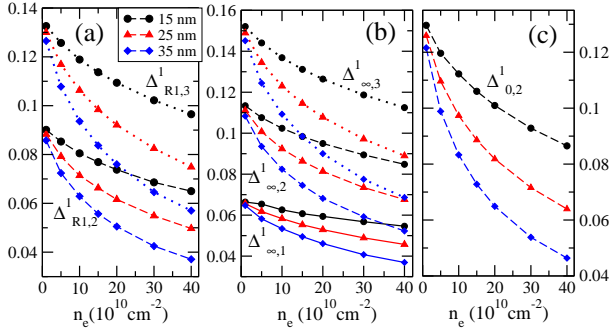}
\caption{(color online) The variation of critical energies corresponding to spin-one modes (Fig.~\ref{Fig_spec_13}(d)) 
with electron densities and width of the quantum wells at filling factor 1/3.
Critical energies in the unit of $e^2/(\epsilon \ell)$ for (a) rotons: $\Delta^1_{R1,2}$, and $\Delta^1_{R1,3}$; (b) high-momentum modes: $\Delta^1_{\infty,1}$,
$\Delta^1_{\infty,2}$, and $\Delta^1_{\infty,3}$; (c) long-wavelength mode: $\Delta^1_{0,2}$.}
\label{Fig_spin_13}
\end{figure}

 Figure \ref{Fig_cm_13} shows the variation of critical energies (Fig.~\ref{Fig_spec_13}), {\it viz.}, $\Delta^0_{R1,1}$, $\Delta^0_{R1,2}$,
 $\Delta^0_{R1,3}$, $\Delta^0_{\infty,1}$, $\Delta^0_{\infty,2}$, $\Delta^0_{\infty,3}$, and $\Delta^0_{0,1}$ of spin-zero modes at $\nu =1/3$
 with electron density $n_e$ at different values of widths, $d$, of quantum wells. The dependence of the critical energies, {\it viz.}, $\Delta^1_{R1,2}$,
 $\Delta^1_{R1,3}$, $\Delta^1_{\infty,1}$, $\Delta^1_{\infty,2}$, $\Delta^1_{\infty,3}$, $\Delta^1_{0,2}$ of spin-one modes shown in Fig.~\ref{Fig_spin_13}
 at $\nu = 1/3$ on $n_e$
 and $d$. All these energies decreases with the increase of both $n_e$ and $d$, as the effective Coulomb repulsion at short distances decreases 
 (Fig.~\ref{fig.LDA_pot}).


\begin{figure}[h]
\vspace{0.5cm}\hspace{0.5cm}
\includegraphics[angle=270,width=6.0cm]{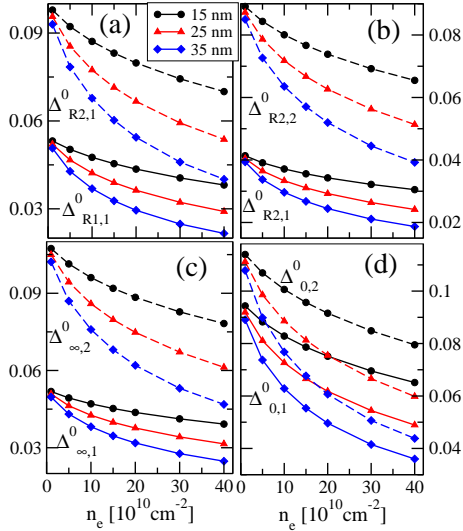}
\caption{(color online) The variation of critical energies corresponding to spin-zero modes (Fig.~\ref{Fig_spec_25}(b)) 
with electron densities and width of the quantum wells at filling factor 2/5.
Critical energies in the unit of $e^2/(\epsilon \ell)$ for (a) rotons: $\Delta^0_{R1,1}$, $\Delta^0_{R2,1}$; (b) rotons: $\Delta^0_{R1,2}$, 
$\Delta^0_{R2,2}$; (c) high-momentum modes: $\Delta^0_{\infty,1}$ and
$\Delta^0_{\infty,2}$; (d) long-wavelength modes: $\Delta^0_{0,1}$ and $\Delta^0_{0,2}$.}
\label{Fig_cm_25}
\end{figure}

\begin{figure}[h]
\vspace{1.0cm}\hspace{0.5cm}
\includegraphics[angle=270,width=8.0cm]{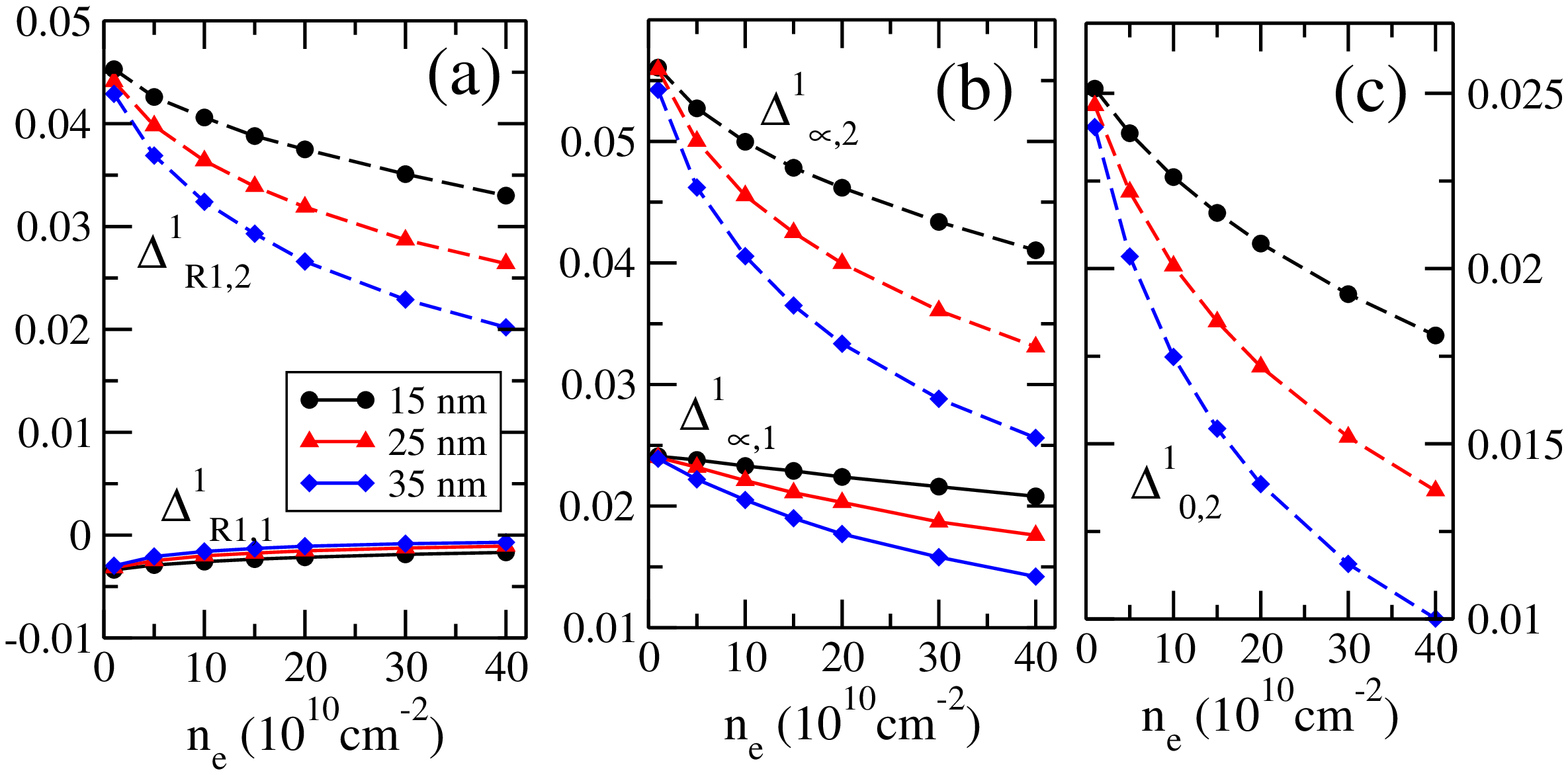}
\caption{(color online) The variation of critical energies corresponding to spin-one modes (Fig.~\ref{Fig_spec_25}(d)) 
with electron densities and width of the quantum wells at filling factor 2/5.
Critical energies in the unit of $e^2/(\epsilon \ell)$ for (a) rotons: $\Delta^1_{R1,1}$ and $\Delta^1_{R1,2}$; 
(b) high-momentum modes: $\Delta^1_{\infty,1}$ and
$\Delta^1_{\infty,2}$; (d) long-wavelength mode: $\Delta^1_{0,2}$.}
\label{Fig_spin_25}
\end{figure}

The dependence of critical energies of spin-zero and spin-one modes (marked in Fig.~\ref{Fig_spec_25}) at $\nu=2/5$ on $n_e$ and $d$ are shown respectively in
Figs.~\ref{Fig_cm_25} and \ref{Fig_spin_25}. The energy of $\Delta^1_{R1,1}$ is negative at all $n_e$ and $d$ considered in this paper. The magnitude of all
the energies decrease with the increase of $n_e$ and $d$.

\begin{figure}[h]
\vspace{0.5cm}\hspace{0.5cm}
\includegraphics[angle=270,width=6.7cm]{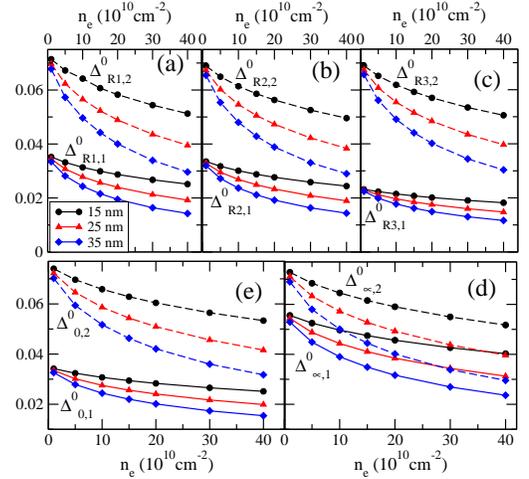}
\caption{(color online) The variation of critical energies corresponding to spin-zero modes (Fig.~\ref{Fig_spec_37}(b)) 
with electron densities and width of the quantum wells at filling factor 3/7.
Critical energies in the unit of $e^2/(\epsilon \ell)$ for (a) rotons: $\Delta^0_{R1,1}$, $\Delta^0_{R1,2}$; (b) rotons: $\Delta^0_{R2,1}$, 
$\Delta^0_{R2,2}$; (c) rotons: $\Delta^0_{R3,1}$, $\Delta^0_{R3,3}$; (d) high-momentum modes: $\Delta^0_{\infty,1}$ and
$\Delta^0_{\infty,2}$; (e) long-wavelength modes: $\Delta^0_{0,1}$ and $\Delta^0_{0,2}$.}
\label{Fig_cm_37}
\end{figure}

\begin{figure}[h]
\vspace{0.5cm}\hspace{0.5cm}
\includegraphics[angle=270,width=8.0cm]{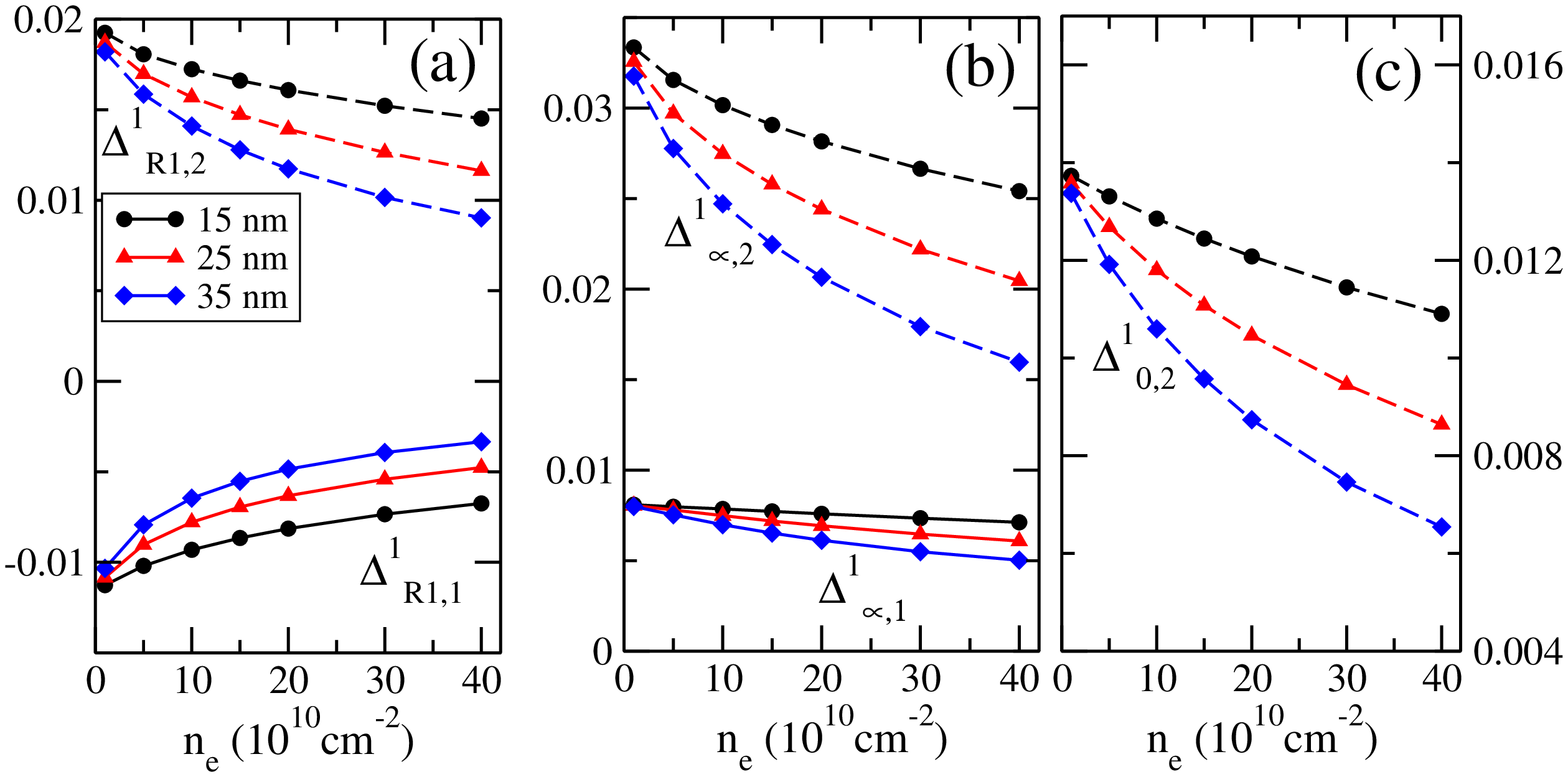}
\caption{(color online) The variation of critical energies corresponding to spin-one modes (Fig.~\ref{Fig_spec_37}(d)) 
with electron densities and width of the quantum wells at filling factor 3/7.
Critical energies in the unit of $e^2/(\epsilon \ell)$ for (a) rotons: $\Delta^1_{R1,1}$ and $\Delta^1_{R1,2}$; 
(b) high-momentum modes: $\Delta^1_{\infty,1}$ and
$\Delta^1_{\infty,2}$; (c) long-wavelength mode: $\Delta^1_{0,2}$.}
\label{Fig_spin_37}
\end{figure}

\begin{figure}[h]
\vspace{0.5cm}\hspace{0.5cm}
\includegraphics[angle=270,width=6.0cm]{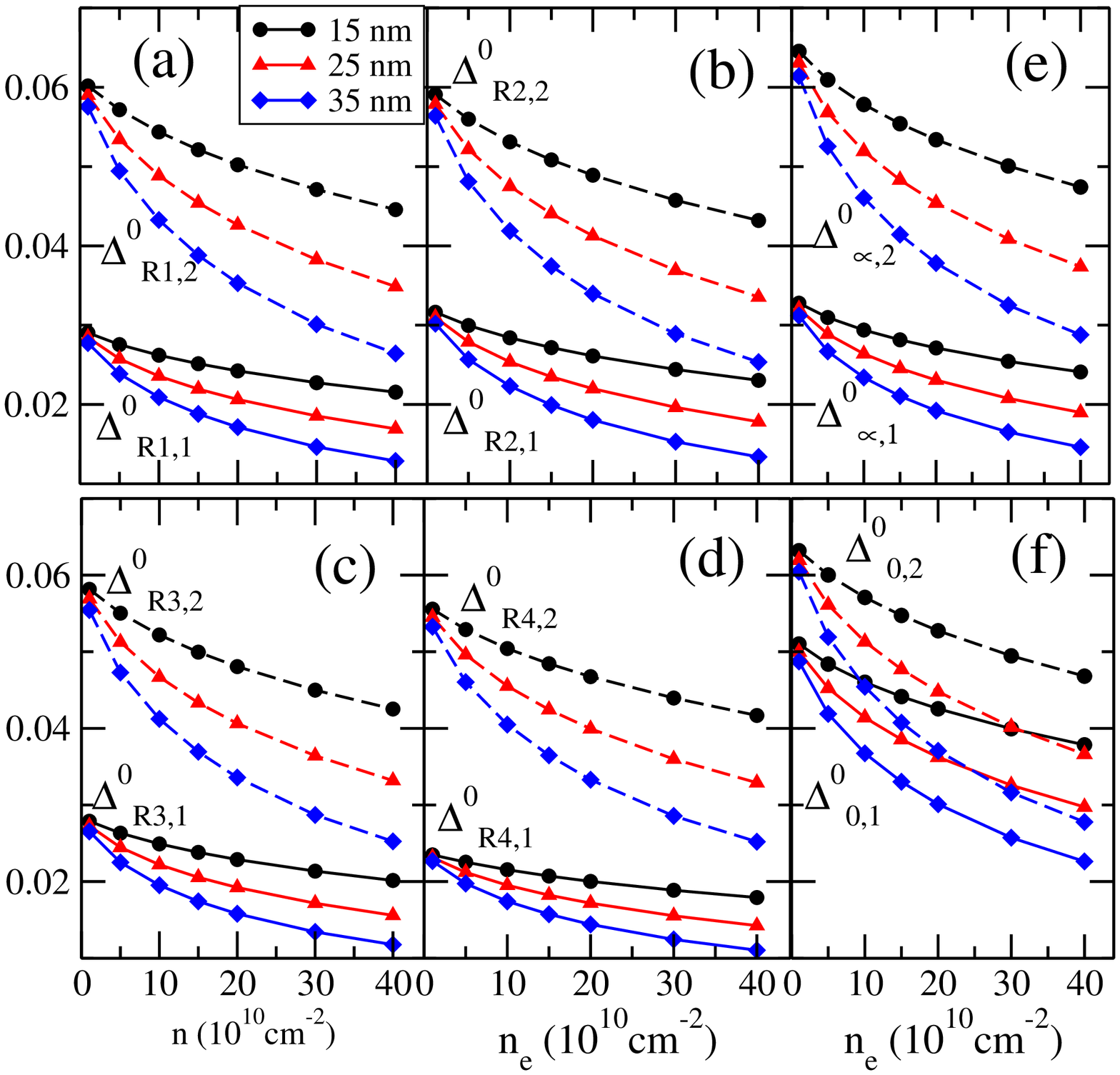}
\caption{(color online) The variation of critical energies corresponding to spin-zero modes (Fig.~\ref{Fig_spec_49}(b)) 
with electron densities and width of the quantum wells at filling factor 4/9.
Critical energies in the unit of $e^2/(\epsilon \ell)$ for (a) rotons: $\Delta^0_{R1,1}$, $\Delta^0_{R1,2}$; (b) rotons: $\Delta^0_{R2,1}$, 
$\Delta^0_{R2,2}$; (c) rotons: $\Delta^0_{R3,1}$, $\Delta^0_{R3,2}$; (d) rotons: $\Delta^0_{R4,1}$, $\Delta^0_{R4,2}$; 
(e) high-momentum modes: $\Delta^0_{\infty,1}$ and
$\Delta^0_{\infty,2}$; (e) long-wavelength modes: $\Delta^0_{0,1}$ and $\Delta^0_{0,2}$.}
\label{Fig_cm_49}
\end{figure}

\begin{figure}[h]
\vspace{0.5cm}\hspace{0.5cm}
\includegraphics[angle=270,width=6.0cm]{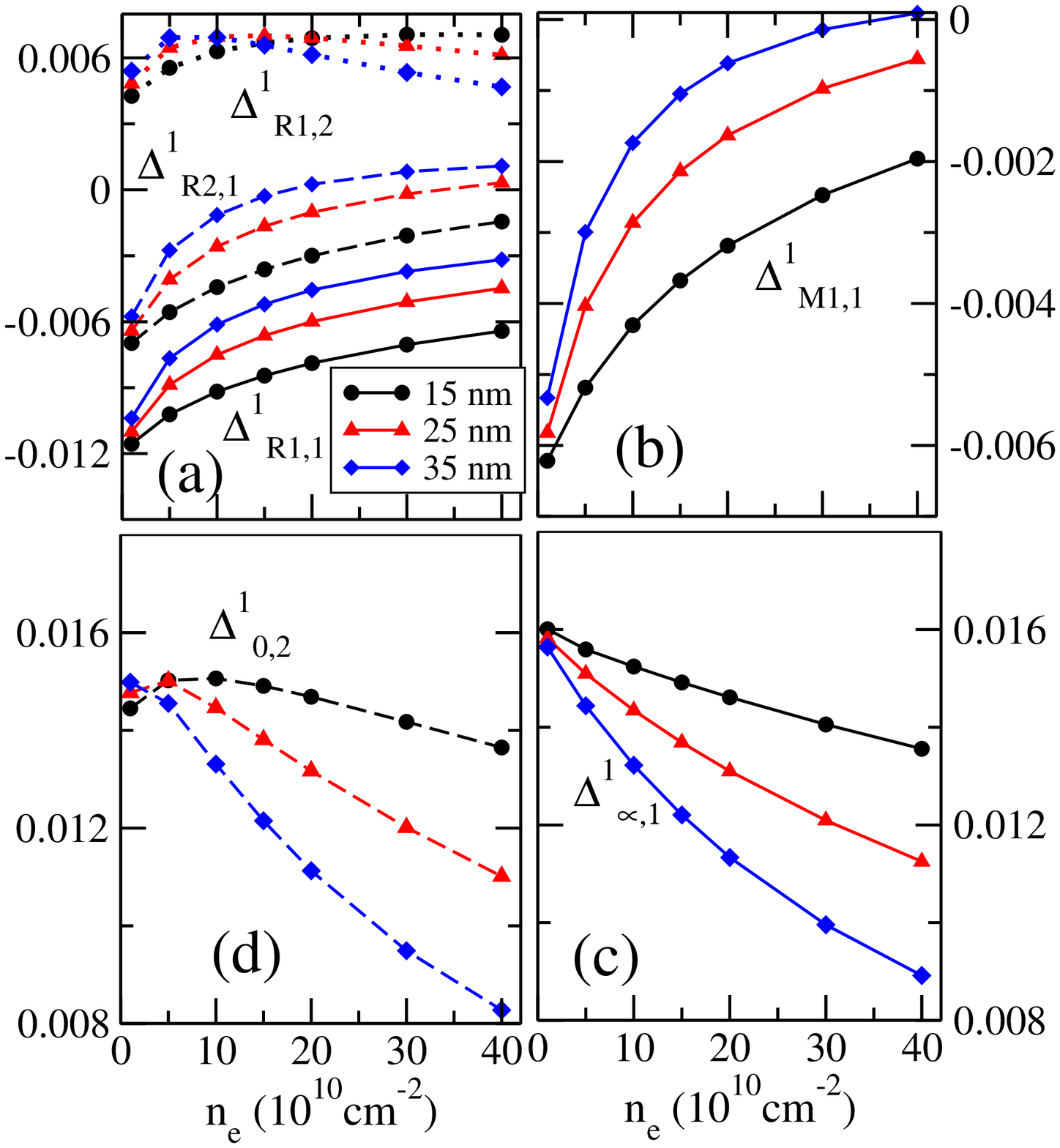}
\caption{(color online) The variation of critical energies corresponding to spin-one modes (Fig.~\ref{Fig_spec_49}(d)) 
with electron densities and width of the quantum wells at filling factor 4/9.
Critical energies in the unit of $e^2/(\epsilon \ell)$ for (a) rotons: $\Delta^1_{R1,1}$, $\Delta^1_{R1,2}$ and $\Delta^2_{R3,1}$;
(b) maxon: $\Delta^1_{M1,1}$;
(c) high-momentum mode: $\Delta^1_{\infty,1}$; (d) long-wavelength mode: $\Delta^1_{0,2}$.}
\label{Fig_spin_49}
\end{figure}

Figures \ref{Fig_cm_37} and \ref{Fig_spin_37} respectively show critical energies (marked in Fig.~\ref{Fig_spec_37}) of spin-zero and spin-one modes at $\nu =3/7$.
As for $\nu=2/5$, the energy $\Delta^1_{R1,1}$ is negative for all densities and widths. The critical energies (marked in Fig.~\ref{Fig_spec_49}) at $\nu =4/9$
for spin-zero and spin-one modes are shown in Figs.~\ref{Fig_cm_49} and \ref{Fig_spin_49} respectively as they depend on $n_e$ and $d$.



\section{Conclusion}

We have considered finite thickness corrections to the excitation energies, but the other important parameters like disorder and Landau level mixing
have not been considered for estimating the excitation energies because of the unavailability of suitable tools in treating those parameters within the
technique that have been employed here. It has been observed that the actual excitation energy could be up to 50$\%$ smaller \cite{Kukushkin09} than the energy estimated
using finite-thickness correction only.

Some of the modes obtained here have already been observed in various experiments. The lowest spin-zero modes at $\nu =2/5$, 3/7, and 4/9 have been observed in a surface
acoustic wave (SAW) experiment \cite{Kukushkin09}. Inelastic light scattering experiments determine the critical energies at which density of states become large due to vanishingly small
slope in the energy dispersions. Unlike SAW experiments, ILS experiments cannot determine the full dispersion but the latter has advantage over the former in determining 
excitaions at higher energies and also the spin-one modes. The modes that have been observed \cite{Pinczuk93,Kang01,Dujovne03,Dujovne05,Majumder11_2,Majumder11_1}
so far in ILS experiments have been interpreted as $\Delta^0_{0,1}$,
$\Delta^0_{R1,1}$, $\Delta^0_{\infty ,1}$, $\Delta^0_{R1,2}$, $\Delta^0_{\infty ,2}$, $\Delta^0_{R1,3}$, $\Delta^0_{\infty ,3}$, $E_z$, 
$\Delta^1_{\infty,1}$, $\Delta^1_{0,2}$, $\Delta^1_{R1,2}$, $\Delta^1_{0,3}$, and $\Delta^1_{R1,3}$ at $\nu =1/3$; $\Delta^0_{0,1}$, $\Delta^0_{R1,1}$, $\Delta^0_{R2,1}$, $\Delta^0_{\infty ,1}$,
$E_z$, $\Delta^1_{R1,1}$, $\Delta^1_{\infty,1}$ at $\nu =2/5$; $E_z$, $\Delta^1_{R1,1}$ and $\Delta^1_{\infty,1}$ at $\nu = 3/7$; and $E_z$, $\Delta^1_{R1,1}$,
$\Delta^1_{M1,1}$, and $\Delta^1_{R2,1}$, and $\Delta^1_{\infty,1}$ at $\nu=4/9$. The mode-splitting at long-wavelength for $\nu =1/3$ has also been reported in ILS
experiments.


We hope that the results presented here will stimulate further inelastic light scattering experiments to observe the critical energies
for the higher energy spin-zero and spin-one modes in the fully polarized FQHE states at the filling factors 2/5, 3/7, and 4/9.
The calculations using SMA \cite{Girvin85,Park00} as well as level-1$^+$ CF excitons \cite{Dev92,JK,Scarola00} suggest that the lowest
spin-zero mode at $\nu =n/(2n+1)$ will have $n$-rotons. This has been experimentally verified as well in a SAW experiment \cite{Kukushkin09}.
We here predict that the next higher spin-zero modes will also have $n$-roton minima. It will be interesting if such an observation is also made
in a future SAW experiment. The real difficulty in such an experiment is to probe excitations at higher energy. This is the reason for not
observing dispersion at $\nu =1/3$ in Ref.~\onlinecite{Kukushkin09}. Nonetheless, it is possible to detect the second spin-zero mode (Fig.~\ref{Fig_spec_49}(b))
at $\nu =4/9$ as its energy is in the same ball park of the lowest mode at $\nu = 2/5$ which has already been found in SAW experiment.
However the high-energy excitations should, in principle, be accessible to the experiments like SAW and time domain capacitance spectroscopy \cite{Dial10}. 

\section*{Acknowledgment}

 We are grateful to J. K. Jain for stimulating and fruitful discussions.

\end{document}